\documentclass[%
 reprint,
 amsmath,amssymb,
 aps,
]{revtex4-2}

\usepackage{graphicx}
\usepackage{dcolumn}
\usepackage{bm}
\usepackage{hyperref}
\begin{document}
\title{Rigorous proof of the Strutinsky energy theorem and foundations of nuclear density functional theory}
\author{Chong Qi}
 \affiliation{Department of Physics, KTH Royal Institute of Technology, AlbaNova University Centre, 106 91 Stockholm, Sweden}

\date{\today}

\begin{abstract}
We have derived a rigorous theoretical proof of the Strutinsky energy theorem. This proof not only provides a proper interpretation of the shell-correction decomposition, resolving decades of confusion, but also lays a foundation for constructing nuclear density functionals.
\end{abstract}

\maketitle

The shell structure has remained the central organizing principle of nuclear physics for nearly six decades since its formulation. In microscopic terms, shell structure is encoded in the single-particle energy spectrum, which can be generated either from phenomenological single-particle potentials, such as Woods–Saxon or Nilsson-type models, or be derived self-consistently from an underlying two-body interaction within mean-field or general many-body frameworks. However, a fundamental limitation arises from the fact that nucleons generate their own mean field through mutual interactions. It is thus crucial to recognize that one cannot simply sum the occupied single-particle energies to obtain the total binding energy of the nucleus, which inevitably leads to double counting of interaction contributions, reflecting the fact that the independent-particle picture is only an effective representation of a correlated many-body system.

A decisive breakthrough in the quantitative description of nuclear shell effects was achieved by V. M. Strutinsky, who introduced the shell-correction method \cite{Strutinsky1967,Strutinsky1968}. The remarkable success of this approach lies in its ability to combine macroscopic bulk systematics with microscopic single-particle structure, thereby reproducing binding energies, equilibrium deformations, and fission barriers across the nuclear chart \cite{NILSSON19691,Brack1972}. For decades, the Strutinsky method has been regarded as the gold standard of the macroscopic–microscopic framework and remains, even today, among the most precise and practical tools for global nuclear structure calculations \cite{SOBICZEWSKI2007292,Moller2012,wang2014surface,wu2015global}. 

In this work, we revisit a fundamental aspect that has remained largely unexamined: the theoretical foundation of the Strutinsky theorem itself and the proper interpretation of the shell-correction procedure. Traditionally, the method has been described within the macroscopic–microscopic paradigm, where the shell correction is interpreted as the difference between the discrete sum of single-particle energies and a smoothed contribution associated with an averaged level density. We will demonstrate, however, that this conventional interpretation is conceptually flawed. In particular, the commonly invoked “mean level density” does not possess an independent physical meaning. Our analysis provides a rigorous proof of the underlying theorem and establishes a new interpretation of the shell-correction method, clarifying its formal structure and resolving long-standing conceptual ambiguities.

The standard textbook interpretation of the Strutinsky shell-correction method decomposes the total nuclear energy into two distinct terms: a smooth, macroscopic component $\tilde E^{(0)}$ and an oscillatory shell contribution in the form 
$\delta E_{\mathrm{shell}}$, so that
\begin{equation}\label{SET}
E = \tilde E^{(0)} + \delta E^{(1)}_{\mathrm{shell}} + \delta E^{(2)},
\end{equation}
where $\delta E^{(2)}$ represents a presumed small, general second-order contribution.
Introducing the smoothed level density
$\tilde g(\varepsilon)$,
the smoothed energy is defined as
\begin{equation}
\tilde E_{\mathrm{sp}} 
= \int_{-\infty}^{\tilde{\varepsilon}_F} 
\varepsilon\, \tilde g(\varepsilon)\, d\varepsilon,
\qquad
N = \int_{-\infty}^{\tilde{\varepsilon}_F} 
\tilde g(\varepsilon)\, d\varepsilon,
\end{equation}
where $N$ is the total number of protons or neutrons. Here and in the following, quantities with a tilde denote smoothed (averaged) counterparts of the corresponding microscopic quantities. In particular, 
$\tilde g(\varepsilon)$ and $\tilde{\varepsilon}_F$ are obtained from a smoothing procedure applied to the discrete spectrum $g(\varepsilon)$ and Fermi energy $\varepsilon_F$, respectively.
The shell correction then reads
\begin{equation}
\delta E^{(1)}_{\mathrm{shell}}
= E_{\mathrm{sp}} - \tilde E_{\mathrm{sp}}.
\end{equation}
The shell structure of nuclei is encoded in the single-particle spectrum 
$\{\varepsilon_i\}$ generated by an effective mean-field potential. The naive sum of occupied single-particle energies gives
\begin{equation}
E_{\mathrm{sp}} = \sum_{i=1}^{N} \varepsilon_i 
= \int_{-\infty}^{\varepsilon_F} \varepsilon\, g(\varepsilon)\, d\varepsilon,
\end{equation}
which, as mentioned above, does not represent the total binding energy due to double counting inherent 
to the two-body origin of the mean field.

In practical implementations, the average single-particle level density entering the shell-correction method is constructed using the smoothing prescription introduced by Strutinsky himself. It is based on a curvature-corrected convolution with a Gaussian kernel that ensures plateau stability and removes spurious oscillations.  The averaged level density obtained through this method does not arise from a systematic semiclassical expansion of the many-body functional. Instead, it emerges from an optimized filtering procedure applied directly to the discrete spectrum—functioning effectively as an optimized ``black box". Contrary to the idealized smooth curves often depicted in textbooks (see Fig.~2.28 in Ref.~\cite{Ring1980}), this resulting density frequently retains non-trivial fluctuations. Consequently, there is no formal guarantee that such a density will remain thermodynamically or structurally compatible with the smooth macroscopic energy $\tilde{E}$, which is de facto identified with the liquid-drop model.
At the same time, certain limitations are well known: the method depends on intrinsic smoothing parameters, requires careful plateau-condition analysis, and faces conceptual and technical difficulties in the treatment of the particle continuum.

Numerous alternative approaches have been proposed to evaluate the average level density within semiclassical theory and extended Thomas–Fermi methods \cite{GROSS197241,Brack1973,BRACK1975421,Bhagwat2023,Ring1980,Pomorski2004,Aboussir1995,Jennings1973,Jennings1975,Jennings1975a,Badhuri1971,Vertse1998}.
From this semiclassical perspective, shell effects arise as quantum oscillations superimposed on a smooth background determined by classical dynamics and asymptotic limits. While these procedures clarify the need for oscillatory corrections and show that the Strutinsky method is a leading-order approximation to the semiclassical treatment, they operate  at the level of the single-particle density of states only and do not establish a variational decomposition of the total energy into well-defined smooth and first-order terms.

The central issue, therefore, is not the technical construction of a smoothed level density. Rather, it is the absence of a formal density-functional theorem establishing how the total many-body energy separates into smooth and oscillatory components. Without such a theorem, the identification of shell corrections with fluctuations of the level density remains interpretative rather than derivational.
Strutinsky attempted to justify Eq.~(\ref{SET}),  often called the Strutinsky energy theorem, within the framework of self-consistent Hartree–Fock theory~\cite{Strutinsky1968,Bunatian1972,Brack1972}. The central idea was the following. In Hartree–Fock, the total energy of a finite Fermi system can be expressed as a functional of occupied single-particle orbitals.
If the system is governed by a two-body potential $V_{ij}$, the Hamiltonian is:
$$H = \sum_i t_i + \sum_{i<j} V_{ij}.$$
Within Hartree-Fock, one can use the approximation of
\begin{equation}
    E = \sum \langle t_i \rangle + \frac{1}{2}\langle V \rangle=\sum \epsilon_i - \frac{1}{2}\langle V \rangle
\end{equation}
 to correct for the double-counting of the two-body potential. Here $\epsilon_i$ is the Hartree-Fock eigenvalue, determined by the corresponding Hartree-Fock potential $U(r)$, and $\langle V \rangle$ is the total expectation value of the two-body force. Nevertheless, in the context of a non-self-consistent model like the Woods-Saxon, we cannot calculate the $\langle V \rangle$ term from the potential itself because the potential is phenomenological and does not explicitly contain the two-body interaction.

Assuming there is an averaged density $\tilde\rho $ with a corresponding smooth potential $\tilde U$, the total energy can be approximated as (c.f., Eq.~(18) in Ref. \cite{Strutinsky1968})
\begin{equation}
    E \approx  \sum \tilde\epsilon_i - \frac{1}{2}\langle \tilde V \rangle_{\tilde\rho } + O\left[(\delta \rho)^{2}\right],
\end{equation}
where \(
\delta\rho = \rho - \tilde\rho,
\delta U = U - \tilde U\) and to first order,
\begin{equation}
\epsilon_i
=
\tilde\epsilon_i
+
\langle  \delta U \rangle
+
O\left[(\delta \rho)^{2}\right].
\end{equation}
The same line of reasoning was subsequently adopted in the standard textbooks of Bohr and Mottelson~\cite{Bohr1998} [Eq.~(6-101)] and Ring and Schuck~\cite{Ring1980} and by Bethe \cite{Bethe1971} [Eq.~(7)]. Brack and collaborators reformulated it as a smoothed Hartree-Fock orbital occupation \cite{BRACK1975421,Brack1997}.

Strutinsky explicitly acknowledged that the resulting expression may not be of practical use  for fully self-consistent calculations but does indicate that one can separate a rapidly oscillating contribution contained in the single-particle sum from a smooth term depending on $\tilde\rho$. The argument, however, is not a formal proof of the shell-correction method but rather a perturbative reorganization of Hartree–Fock framework: the shell correction emerges as the fluctuating part of the single-particle energy sum but it doesn't provide a way of subtracting a suitably smoothed average, while it is unclear what the smooth component  contains. The success of the method thus rests on the physical plausibility and numerical stability of this separation, rather than on a fully controlled functional derivation. In other words, the Hartree–Fock decomposition does not guarantee that the smooth component obtained by spectral averaging is physically meaningful or corresponds to the stationary point of any variational principle.

A further source of ambiguity arises from the identification of the single-particle energies $\{\tilde\epsilon_i\}$ with the eigenvalues of an averaged (non–self-consistent) potential (see discussions on page 371, Vol. II Ref. \cite{Bohr1998} and page 20 in Ref. \cite{Strutinsky1968}). These eigenvalues are, in principle, distinct from those obtained in the original self-consistent solution, so that the separation between smooth and fluctuating parts involves a replacement of the underlying spectrum rather than a transformation within the same self-consistent framework. A long-neglected perspective is that, if this interpretation is taken at face value, a fundamental inconsistency emerges. Even when the total energies obtained in this manner are numerically similar, the underlying theoretical frameworks remain conceptually distinct. The Hartree–Fock scheme and the phenomenological single-particle model embedded in the Strutinsky method are based on different Hamiltonians and therefore yield different single-particle eigenvalues and eigenstates and are two inequivalent mean-field descriptions. This dilemma was actually implied in the statement of Strutinsky and collaborators in Ref. \cite{Bunatian1972}: ``$\tilde \rho$ and $\tilde U$ are not self-consistent and they do not correspond to any real physical
system. However the potential $\tilde U$ may in principle be identified with phenomenological
non-consistent potentials of the shell model".

These limitations do not diminish the practical success of the shell-correction method. Rather, they indicate that the original justification—while physically insightful—falls short of a well-defined theorem. The smoothing of the level density functions as an effective computational device. At the same time, its connection to a fundamental decomposition of the total energy remains conceptually incomplete.

In the following, we present for the first time a formal proof based on density functional theory (DFT). Instead of modifying the single-particle spectrum within Hartree–Fock, we derive directly at the level of the total energy functional a decomposition into a smooth density-dependent part and a controlled first-order correction around a reference density. This approach preserves variational consistency and does not rely on external spectral smoothing, thereby placing the shell-correction concept on a well-defined functional foundation.

DFT provides a preeminent theoretical framework for describing complex quantum many-body systems. Its core innovation lies in shifting the focus from the intractable, full many-body wavefunction to the spatial density, $\rho(\mathbf{r})$, thereby simplifying the analysis while capturing the essential physics of the system. Although a strictly well-defined DFT for nuclear systems remains elusive due to the immense complexity of the nucleon-nucleon interaction, the DFT philosophy has been successfully extended to nuclear physics. 
Consider a system of fermions interacting through a local two-body potential
\(
V(\mathbf r_1,\mathbf r_2),
\)
the ground state energy $E$, which is a unique functional of the density $\rho(\mathbf{r})$ within the DFT context, can be written as
\begin{eqnarray}\label{DFT}
E[\rho] =
T[\rho]
+ \frac{1}{2}
\int d^3r\, d^3r'\,
V(\mathbf r,\mathbf r')\,
\rho(\mathbf r)\rho(\mathbf r')\nonumber  \\=\sum \varepsilon_{i}
- \frac{1}{2}
\int d^3r\, d^3r'\,
V(\mathbf r,\mathbf r')\,
\rho(\mathbf r)\rho(\mathbf r'),
\end{eqnarray}
where $ T[\rho]=\int t \rho(\mathbf{r}) \, d\mathbf{r}$  is the kinetic energy functional, $\varepsilon_{i}$ denotes the self--consistent single-particle energy, which includes both kinetic and mean-field contributions, and \( \rho(\mathbf r) \) denotes the ground-state
density subject to the particle-number constraint $\int \rho(\mathbf{r})\, d\mathbf{r} = N$. The existence and uniqueness of this functional are guaranteed by the Hohenberg–Kohn theorem \cite{Hohenberg1964} which also provides a variational principle, asserting that the energy functional $E[\rho]$ attains its global minimum only at the exact ground-state density $\rho (\mathbf{r})$.

The practical success of DFT is largely attributed to the Kohn–Sham theorem~\cite{Kohn1965}, which maps the complex, interacting many-body problem onto a system of non-interacting particles moving within an effective mean-field potential. 
Under the assumption that the density is represented by a set of single-particle orbitals, $\rho(\mathbf{r}) = \sum_i |\phi_i(\mathbf{r})|^2$, the self-consistent effective Hamiltonian (Kohn-Sham or Hartree-Fock type) is then formally  defined as the functional derivative of the energy with respect to the density:
\begin{eqnarray}h = \frac{\delta E}{\delta \rho} = t + \int V(\mathbf{r}, \mathbf{r}') \rho(\mathbf{r}') \, d\mathbf{r}'.\end{eqnarray}
At the exact ground-state density, the variational principle dictates that this derivative equates to the chemical potential, $\frac{\delta E}{\delta \rho} = \mu$, ensuring that the density is self-consistent with the potential it generates.

Given a reference density $\tilde{\rho}$, the ground-state energy functional $E[\rho]$ can be expanded as a Taylor series:\begin{equation}E[\rho] = E[\tilde{\rho}] + \int \frac{\delta E}{\delta\rho(\mathbf{r})} \Bigg|_{\tilde{\rho}} \bigl(\rho(\mathbf{r}) - \tilde{\rho}(\mathbf{r})\bigr)d^3r + O(\delta \rho^2).\end{equation}The leading-order term, $\tilde{E} = E[\tilde{\rho}]$, represents the energy evaluated at the reference density which takes the same form as Eq.~(\ref{DFT})
\begin{equation}\tilde{E} = E[\tilde{\rho}]=\int t \tilde{\rho}(\mathbf{r}) d^3r + \frac{1}{2} \int \int V(\mathbf{r},\mathbf{r}') \tilde{\rho}(\mathbf{r})\tilde{\rho}(\mathbf{r}') d^3r d^3r'.\end{equation}
It is important to note that for higher-order terms to be negligible, $\tilde{\rho}$ cannot be merely an arbitrary background; it must be a variational density that remains in the neighborhood of the true ground-state density $\rho$. Conversely, $\tilde{E}$ and $\tilde{\rho}$ do not need to represent the physical ground-state energy and density, or even a "real" physical system. They serve as a baseline for the perturbation, with $\tilde{E}$ being the energy of a system whose density is constrained to the reference distribution $\tilde{\rho}$. Importantly, if $\tilde{\rho}$ is chosen to be a ``smooth" macroscopic density, $\tilde{E}$ effectively captures the bulk properties of the system, such as the liquid-drop energy in nuclei. The linear correction term, $\delta E^{(1)}$, then accounts for the energy shift due to the deviation of the true density $\rho$ from the reference $\tilde{\rho}$.  This term is governed by the effective single-particle Hamiltonian $\tilde{h}$, evaluated at the reference density:
\begin{eqnarray}
\delta E^{(1)} =
\int d^3r\;
\frac{\delta E}{\delta\rho(\mathbf r)}
\Bigg|_{\tilde{\rho}}
\bigl(\rho(\mathbf r) - \tilde{\rho}(\mathbf r)\bigr) = \int \tilde{h} \delta \rho(\mathbf{r}) \, d\mathbf{r},
\end{eqnarray}
where $\rho = \tilde{\rho} + \delta \rho$ and 
\begin{eqnarray}\label{mean-field}
    \tilde{h} = \frac{\delta E}{\delta \rho} \big|_{\tilde{\rho}}.
\end{eqnarray}
Here, $\tilde{h}$ acts as the ``mean field" generated by the reference density. In the context of the shell-correction method, this term provides the link between the macroscopic bulk energy and the microscopic fluctuations arising from the shell structure.

The expansion can be rewritten in the Strutinsky energy theorem form as
\begin{eqnarray}E[\rho] = {E}[\tilde{\rho}] + \delta E^{(1)} + O(\delta \rho^2).
\end{eqnarray}
Here, \( {E}[\tilde{\rho}] \) is presumed to be a smooth functional of the smoothed density \( \tilde{\rho} \).
The smooth density 
$\tilde{\rho}$
is not necessarily uniquely defined; it should be understood as a variationally admissible reference density that filters out shell oscillations while remaining close to the self-consistent solution.
 The term \( \delta E^{(1)} \) represents the first-order correction, which captures all oscillatory energy contributions arising from fluctuations of the density about \( \tilde{\rho} \), and consequently from the discrete nature of the single-particle spectrum from $\tilde{h}$. The remaining term \( O(\delta \rho^2) \) collects second-order and higher-order contributions in the density fluctuations and is generally small.
Already at this stage, it can be stated that this simple expansion provides a formal mathematical foundation for evaluating the shell correction. It establishes a formal link between a self-consistent DFT calculation, utilizing an explicit two-body interaction $V$,and the phenomenological shell model, provided $\tilde{h}$ is replaced by an appropriate external potential. Nevertheless, similar to the challenges encountered in the original Strutinsky formulation, the expansion is not immediately useful in practice, as neither the exact density $\rho$ nor the fluctuation $\delta \rho$ is known a priori.

Let us assume that the reference density is not a randomly valued function but is a generated by a reference set of single-particle orbitals $\{\phi_i^0\}$
\begin{equation}
\tilde\rho (\mathbf{r}) = \sum_i |\phi_i^0(\mathbf{r})|^2.
\end{equation}
From $\tilde\rho$ we construct the Kohn-Sham Hamiltonian in Eq.~(\ref{mean-field}):
\begin{equation}
\tilde{h}[\tilde\rho] = -\frac{1}{2}\nabla^2 + \tilde{U}(r)[\tilde\rho](\mathbf{r})
\end{equation}
where $\tilde{U}(r)$ is the effective single-particle potential generated by the density $\tilde\rho$.
Solving $\tilde{h}$ yields a new set of eigenstates and eigenvalues:
\begin{equation}
\tilde{h}[\tilde\rho]|\phi_i^1\rangle = \varepsilon_i^1|\phi_i^1\rangle
\end{equation}
with updated density:
\begin{equation}
\rho^1(\mathbf{r}) = \sum_i |\phi_i^1(\mathbf{r})|^2.
\end{equation}
Again, the key assumption is that $\tilde\rho$ is a reasonable approximation 
to the true ground state density, so that:
\begin{equation}
\delta\rho(\mathbf{r}) = \rho^1(\mathbf{r}) - \tilde\rho(\mathbf{r}) = \sum_i\left[|\phi_i^1(\mathbf{r})|^2 - |\phi_i^0(\mathbf{r})|^2\right]
\end{equation}
is small. We are, however, generally not at self-consistency: $\phi_i^0$ are 
{not} eigenstates of $\tilde{h}[\tilde{\rho}]$ in general. 
Self-consistency is recovered only when $\phi_i^1 = \phi_i^0$ and 
$\rho^1 = \tilde{\rho}=\rho$.

Substituting the density variation into the first order term:
\begin{eqnarray}
        \delta E^{(1)} = \int \tilde{h}[\tilde{\rho}]\,\delta\rho(\mathbf{r})\,d\mathbf{r}\nonumber\\ 
= \sum_i\int \tilde{h}[\tilde{\rho}]
\left[\phi_i^{1*}(\mathbf{r})\phi_i^1(\mathbf{r}) 
- \phi_i^{0*}(\mathbf{r})\phi_i^0(\mathbf{r})\right]d\mathbf{r}
\end{eqnarray}
which gives:
\begin{equation}
\delta E^{(1)} = \sum_i\left[
\langle\phi_i^1|\tilde{h}[\tilde{\rho}]|\phi_i^1\rangle 
- \langle\phi_i^0|\tilde{h}[\tilde{\rho}]|\phi_i^0\rangle
\right].
\end{equation}
The first term gives the single-particle energy $\{\varepsilon_i^1\}$ by construction.
To estimate $\langle\phi_i^0|\tilde{h}[\tilde{\rho}]|\phi_i^0\rangle$, 
one can expand $|\phi_i^0\rangle$ in the complete eigenbasis of $\tilde{h}[\tilde{\rho}]$:
\begin{equation}
|\phi_i^0\rangle = \sum_j c_{ij}|\phi_j^1\rangle, 
\qquad c_{ij} = \langle\phi_j^1|\phi_i^0\rangle,
\end{equation}
which lead to
\begin{equation}
\langle\phi_i^0|\tilde{h}[\tilde{\rho}]|\phi_i^0\rangle 
= \sum_j |c_{ij}|^2\varepsilon_j^1.
\end{equation}
This is a weighted average of eigenvalues $\varepsilon_j^1$. 
With $\tilde{\rho}$ approaching self-consistent solution, one has $\phi_i^0 \approx \phi_i^1$, so 
$|c_{ij}|^2 \approx \delta_{ij}$, which leads to vanishing $\delta E^{(1)}$ value.

One can now rewrite the energy theorem as 
\begin{eqnarray}
E[\rho']
=
\tilde{E}[\tilde{\rho}]
+
\sum \varepsilon'_{i}
-\int \tilde{h}\tilde{ \rho}(\mathbf{r}) \, d\mathbf{r}+
O(\delta \rho^2),
\end{eqnarray}
where $\rho'$, $\varepsilon'_{i}$ and $E[\rho']$ approximate the true density, Kohn–Sham single-particle energy and the total energy, respectively.
This result offers several critical insights.

Mathematically, the expansion proves that the first-order error in the total energy is determined solely by the fidelity of the single-particle spectrum without requiring explicit knowledge of the underlying two-body interaction.

Physically, the theorem rigorously justifies that a potential reproducing the correct single-particle levels can capture the first-order correction to the total energy. This moves beyond the common ``loosely interpreted" textbook characterization of phenomenological potentials as representation of unphysical smooth background. 

Conceptually, the theorem clarifies that the ``smoothness" required is a property of the reference density $\tilde{\rho}$, rather than a smoothing of the level density itself, as is often mischaracterized. By anchoring the separation in a smooth density, the shell energy becomes theoretically sound and physically meaningful to first order.

From a DFT perspective, this framework functions as an approximate formulation where the first-order correction depends only on density fluctuations. Since inconsistencies between the phenomenological potential and the true mean field enter only at the second order ($O(\delta \rho^2)$), a well-chosen phenomenological potential allows $\delta E^{(1)}$ to accurately represent the system's true shell energy.

Self-consistent nuclear structure theory has historically relied instead on Hartree–Fock or Hartree–Fock–Bogoliubov approaches. These methods are  not density functional theory in the strict Hohenberg–Kohn sense. 
The expansion of the total energy around a reference density $\tilde{\rho}(\mathbf r)$ may allow us to build a non-self-consistent approximation of the nuclear DFT without requesting explicit knowledge on the two-body force in the form
\begin{equation}
{E}[\tilde{\rho}]=\int \mathcal{F}_{\theta}\left(\tilde{\rho}(\mathbf{r}), \nabla \tilde{\rho}(\mathbf{r}), \nabla^{2} \tilde{\rho}(\mathbf{r}), \ldots\right) d \mathbf{r}
\end{equation}

One great advantage the approximation provides is that
 $V[\rho]$ encodes all nucleon-nucleon interactions
 \begin{equation}
\tilde{U}[\rho](\mathbf{r}) = \frac{\delta V}{\delta\rho(\mathbf{r})}, 
\qquad
\tilde{h}[\rho] = -\frac{1}{2m}\nabla^2 + \tilde{U}[\rho](\mathbf{r}).
\end{equation}
One can further derive the first-order correction as 
\begin{equation}
\delta E^{(1)} = \sum_i\varepsilon_i^1 - T[\tilde{\rho}] 
- \int \tilde{U}[\tilde{\rho}](\mathbf{r})\tilde{\rho}(\mathbf{r})\,d\mathbf{r}.
\end{equation}
Substituting into $E[\tilde\rho]$, one obtains
\begin{eqnarray}
E[\rho'] = \underbrace{T[\tilde{\rho}] + V[\tilde{\rho}]}_{E[\tilde{\rho}]} 
+ \sum_i\varepsilon_i^1 - T[\tilde{\rho}] 
- \int \tilde{U}[\tilde{\rho}]\tilde{\rho}\,d\mathbf{r}\nonumber\\
= \sum_i\varepsilon_i^1 
+ V[\tilde{\rho}] 
- \int \frac{\delta V}{\delta\tilde{\rho}(\mathbf{r})}\tilde{\rho}(\mathbf{r})\,d\mathbf{r}
\end{eqnarray}
where
$T[\tilde{\rho}]$ cancels exactly. That can greatly simplify the searching of the functional.

The present formalism opens several perspectives for practical nuclear structure calculations. The decomposition indicates that phenomenological mean-field potentials can be systematically interpreted as generators of single-particle energies within a density-functional expansion, thereby providing a formal link between macroscopic–microscopic models and DFT. This viewpoint may also guide the construction of fully self-consistent energy density functionals by constraining their single-particle spectra at the level of the first-order correction. Furthermore, the framework can be extended to include pairing correlations within the context of density matrix functional theory, as it provides the single-particle orbitals and energies required for exact pairing treatments within a given configuration space.

To summarize, the Strutinsky shell-correction method  serves as a vital semiclassical bridge between the macroscopic ``liquid-drop" behavior of nuclei and the microscopic ``quantum shell" effects in nuclear physics. Nevertheless, the theoretical foundation of this decomposition has never been placed on fully well-defined ground. Early attempts to justify the energy theorem relied on Hartree–Fock arguments which however did not provide a consistent derivation of the  decomposition. As a result, the standard interpretation, classical bulk energy supplemented by quantum shell oscillations extracted from a smoothed level density, rests on conceptual grounds that are, at best, incomplete.
This widespread interpretation has led to a deeper misunderstanding on the nature of the phenomenological shell model and  the subtraction of a smooth single-particle level density.

In this work, we revisit the problem from the standpoint of density functional theory. Rather than starting from single-particle spectra, we begin from the total energy functional itself. We derive a new decomposition theorem in which the total energy functional is separated into a smooth functional part and a well-defined first-order correction. Crucially, this separation is formulated in terms of the \emph{density}, rather than in terms of the long-misunderstood single-particle \emph{level density}. The first-order correction depends on the introduction of a smooth density that satisfies the variational structure of the functional, thereby preserving consistency with the underlying many-body framework.
This perspective resolves longstanding ambiguities in the interpretation of the energy theorem and provides a systematic foundation for extending shell-correction ideas within modern density functional theory.

\bibliography{apssamp}
\end{document}